# COMPOSITON OF TANTALUM NITRIDE THIN FILMS GROWN BY LOW-ENERGY NITROGEN IMPLANTATION: A FACTOR ANALYSIS STUDY OF THE Ta 4f XPS CORE LEVEL

A. Arranz and C. Palacio[*]

Departamento de Física Aplicada, Facultad de Ciencias, C-XII,

Universidad Autónoma de Madrid, Cantoblanco, 28049-Madrid, Spain

(*) Corresponding author: Fax: ++ 34 91 4974949

e-mail: carlos.palacio@uam.es

ABSTRACT

Tantalum nitride thin films have been grown by "in situ" nitrogen implantation of metallic tantalum at room temperature over the energy range of 0.5–5keV. X-ray photoelectron spectroscopy (XPS) and Factor Analysis (FA) have been used to characterise the chemical composition of the films. The number of the different Ta–N phases formed during nitrogen implantation, as well as their spectral shape and concentrations, have been obtained using principal component analysis (PCA) and iterative target transformation factor analysis (ITTFA), without any prior assumptions. According to FA results, the composition of the tantalum nitride films depends on both the ion dose and ion energy, and is mainly formed by a mixture of metallic tantalum, $\beta$-TaN$_{0.05}$, $\gamma$-Ta$_2$N and cubic/hexagonal TaN phases. The kinetics of tantalum nitridation is charaterised by two stages. In a first stage, the formation of $\beta$-TaN$_{0.05}$ species leads to a strong attenuation of the metallic tantalum signal. During the second stage, $\beta$-TaN$_{0.05}$ transforms into $\gamma$-Ta$_2$N and cubic/hexagonal TaN species. For intermediate ion doses, the concentration of $\gamma$-Ta$_2$N reaches a maximum, subsequently decreasing because of its transformation into cubic/hexagonal TaN phases with increasing the ion dose up to saturation. At saturation, the films are mainly composed of a mixture of $\gamma$-Ta$_2$N and cubic/hexagonal TaN phases, but small Ta$^0$ and $\beta$-TaN$_{0.05}$ signals are also observed. They should be attributed to preferential sputtering of nitrogen and/or to the limited thickness of the film. Comparison of the experimental nitrogen concentration with that obtained using TRIDYN simulations suggests that, in addition to nitrogen implantation and atomic mixing, other mechanisms like ion beam enhanced diffusion or the chemical reactivity of the tantalum substrate towards nitrogen, should be also taken into account at higher ion beam energies.

1. INTRODUCTION

The synthesis and characterisation of tantalum nitride thin films have attracted great interest during the last years due to their different technological applications, especially as protective hard coatings and diffusion barriers in copper metallizations[1-14]. Usually, tantalum nitride thin films have been grown by reactive sputtering (RSP)[5-7], chemical vapour deposition (CVD)[8-11], or nitridation of tantalum in ammonia and nitrogen at high temperatures[15]. Moreover, alternative techniques like ion beam assisted deposition (IBAD)[1-4] and nitrogen implantation[12-14,16,17], that allow the growth of the film keeping the substrate at low or moderate temperatures, have been also used recently by several groups.

Several phases have been reported for the Ta-N system[7,15,18]. In a systematic study, of the tantalum nitridation in an atmosphere of ammonia and nitrogen at temperatures above 900K, Terao[15] has identified using x-ray diffraction (XRD) and electron diffraction, up to seven phases with increasing the nitrogen content: cubic $\beta$-TaN$_{0.05}$, hexagonal $\gamma$-Ta$_2$N, hexagonal $\delta$-TaN, hexagonal $\epsilon$-TaN, hexagonal Ta$_5$N$_6$, tetragonal Ta$_4$N$_5$ and tetragonal Ta$_3$N$_5$. Baba et al[1] have grown tantalum nitride thin films by IBAD with nitrogen ions of 2keV. They conclude that the structure of the films changes from a mixture of Ta$_3$N$_5$, TaN and Ta$_2$N phases to a Ta-N solid solution when the [Ta]/[N] transport ratio is increased. Ensinger et al[2] and Zhang et al[3], have shown using IBAD with nitrogen ions of 25 to 70keV and XRD, that the tantalum nitride films formed consist of a metastable cubic TaN phase. On the other hand, Zhou et al[12] and Raole et al[13] have studied the growth of tantalum nitride films by nitrogen implantation at 80 and 30keV, respectively. According to Zhou et al[12], there is a correlation between the types of nitrides formed and the ion dose. Ta$_2$N forms at lower ion doses and TaN starts to grow at higher ion doses. Finally, for the heaviest nitrogen dose (~$5\times10^{17}$ atoms/cm$^2$), XRD analysis shows the presence of β-TaN$_{0.05}$, γ-Ta$_2$N, δ-TaN and ε-TaN phases. Raole et al[13] have found that β-TaN$_{0.05}$, cubic TaN and hexagonal Ta$_5$N$_6$ phases are formed for low nitrogen ion doses, however, the hexagonal γ-Ta$_2$N and TaN phases are only observed for higher ion doses.

The thickness of the tantalum nitride films grown by either of the above mentioned techniques is big enough to allow their "ex situ" characterisation by XRD, in such a way that the natural oxide formed during the transfer of the sample from the preparation chamber to the analysis chamber does not influence so much the analysis carried out. However, the current trend of decreasing dimensions in integrated circuits leads to the reduction of the thickness of tantalum nitride films used as diffusion barriers against copper[3,11]. Therefore, it should be desirable to characterise such ultrathin films "in situ" with a surface sensitive technique. In a previous work[14], we have formed tantalum nitride thin films by low-energy nitrogen implantation at room temperature. Although qualitative results about the composition of the films as a function of the ion dose and ion energy were obtained from X-ray photoelectron spectroscopy (XPS) measurements, neither information of the kinetics of growth nor quantitative information on the Ta-N phases formed during nitrogen implantation was obtained there because of the complexity of the problem and the lack of reference XPS spectra for the different Ta-N phases. In a recent work, the suitability of Factor Analysis (FA) to obtain quantitative information during the synthesis of hafnium nitride thin films by low-energy nitrogen implantation has been reported by Arranz[19].

Therefore, the aim of this work is to obtain the chemical composition of tantalum nitride thin films grown by N$_2^+$ implantation of tantalum substrates at room temperature (RT) over the energy range of 0.5–5keV. To carry out this task, the evolution of the Ta 4f XPS band during nitrogen implantation has been analysed by means of Factor Analysis (FA). The number of Ta-N phases presents in the films formed, as well as their shape and contribution to the experimental XPS spectra have been obtained. In contrast to previous XPS results[1,3,6,8,14,16,17], such a type of study offers a full description of the chemical composition of the ultrathin tantalum nitride films formed during nitrogen implantation of tantalum substrates.

2. EXPERIMENTAL

Tantalum foils of 99.99% purity manufactured by Reframet-Hoboken were used throughout this work. The experiments were carried out in an ultra-high vacuum chamber at a base pressure better than $6\times10^{-10}$ Torr, rising to $2\times10^{-7}$ Torr of $N_2$ during nitrogen implantation. Tantalum substrates were sputter-cleaned "in situ" using a 3keV $Ar^+$ beam rastered over an area of $1\times1cm^2$ to minimize the development of ion-induced roughness until no impurities were detected by Auger electron spectroscopy (AES). For the nitrogen implantation, a differentially pumped extractor type ion gun (SPECS IQE 12/38) was used. The implantation was carried out at room temperature over the energy range of 0.5–5keV using 5N nitrogen, and rastering the ion beam over an area of $1\times1cm^2$. The ion beam current density was in the range 0.2–3 $\mu A/cm^2$ depending on the ion beam energy. It was measured using a Faraday cup near the sample. The angle between the ion beam and the surface normal was 55º. XPS analysis was performed using a hemispherical analyzer (SPECS EA-10 Plus). The pass energy was 15eV giving a constant resolution of 0.9eV. The Ag $3d_{5/2}$ line was used to calibrate binding energies. A twin anode (Mg and Al) X–ray source was operated at a constant power of 300W, using Mg K$\alpha$ radiation.

## 3. RESULTS AND DISCUSSION

Fig. 1 shows the evolution of the Ta 4f core level spectra of a tantalum substrate for different ion doses up to saturation after implantation at room temperature by 1 and 5keV $N_2^+$ ions. The background has been subtracted using a modified Shirley method[20]. Similar spectra to those of Fig. 1 have been also measured for 0.5, 2 and 3keV ion beam energies. The spectra labelled as $Ta^0$ are representative of the clean-sputtered tantalum substrate. As the nitrogen ion dose increases, an attenuation of the metallic $Ta^0$ doublet at ~21.7eV and the growth of a new broad band in the high binding energy side associated with the formation of tantalum nitride are observed. At saturation, this new band shifts to higher binding energies with decreasing the ion beam energy, suggesting compositional differences in the tantalum nitride film that are dependent on the ion beam energy, as discussed in detail elsewhere[14]. To carry out the quantitative analysis of these data, Factor Analysis (FA), has been applied to the set of eighty Ta 4f XPS spectra measured for the different ion doses and ion beam energies used in this work.

FA is a mathematical technique for the analysis of systems where a property can be represented as a linear combination of several variables. The use of FA to evaluate spectroscopic data is well documented in the literature and the mathematical details will not be repeated here[21-23]. In a first step, known as principal component analysis (PCA) the number of principal components, that is, the different Ta–N phases formed during nitrogen implantation, is determined. The application of the IND criterion[21-24] to the Ta 4f set of spectra gives four principal factors. In a second step, known as Target Testing (TT), the data matrix is decomposed into the abstract components during PCA and then they are transformed into the spectra of the pure components and their respective concentrations (relative fractions) related to every experimental spectrum[21-23]. To carry out this task, an iterative target transformation factor analysis (ITTFA) procedure has been applied to obtain the position and shape of the individual pure components without prior assumptions. The whole mathematical details of ITTFA procedure can be found in the work by Fiedor et al[23] and will only be summarized here. First, a "needle" search has been carried out using four principal factors, to obtain the position of the pure components. This "needle" search allows one to probe the true position of the pure components using a very narrow test spectrum (a delta function). Because of every experimental spectrum has $N=210$ data points, $N$ separate tests with $N$ delta functions are carried out, leading to $N$ separate predicted spectra. The position of the pure components corresponds to the point where the test delta spectrum most closely matches the predicted spectrum, that is, to the point where the residual variance (RV) is lower. Usually, this is shown graphically by plotting the target test response, TTR=−RV, as a function of the position of the delta function. In such a type of plot, called a "needle" search, the TTR reach a maximum at the same position where the maximum of every pure component is located.

The "needle" search has been applied to the set of spectra formed by all the Ta 4f spectra measured (80 spectra). Continuous line in Fig. 2 shows the TTR profile as a function of the binding energy for the full Ta 4f set of spectra. The TTR profile shows multiple maxima due to the severe overlapping between the two maxima (Ta $4f_{7/2}$ and $4f_{5/2}$ peaks) of every principal factor found using the IND criterion. In order to identify the individual components relate to the maxima, an additional set of spectra has been constructed. In this new set, the metallic $Ta^0$ spectrum was subtracted from the experimental spectra by minimizing the difference spectra in the energy range of 21.7−20eV. The "needle" search was also applied to these modified set, and the TTR profile is also shown in Fig. 2 (dashed line). As can be observed, the maxima corresponding to the metallic component at ~21.7eV (peak A) and ~23.6eV (peak A′) (spin-orbit splitting, *sos*=1.9eV) are suppressed allowing an easier identification of the maxima corresponding to the other Ta–N phases. It should be observed that the remaining maxima are enhanced after elimination of those corresponding to the metallic peaks. The peaks B and C at ~22.3 and 22.9eV, respectively, should be attributed to the Ta $4f_{7/2}$ peaks of two different Ta–N phases, and the high binding energy peak D′ at ~25.6eV should correspond to the Ta $4f_{5/2}$ peak of the third Ta–N phase. The TTR profiles of Fig. 2 have been shifted vertically for comparison. It should be pointed out, that the additional maxima observed in the middle region of the TTR profile between peaks C and D′ correspond to the overlapping of the other doublet peaks, B′, C′ and D, associated with B, C and D′, respectively. Anyway, only a maximum of every Ta 4f doublet is necessary to obtain the full shape of a pure component, as will be discussed below.

Once the positions of the four principal components have been obtained by the "needle" search, an iterative target testing process is carried out using delta functions centred at those positions to calculate the initial predicted spectra. The predicted spectra are then used as new test spectra in an iterative way, until the difference between the test and the predicted spectra is within the experimental error[19,23]. From the TTR profiles of Fig. 2, initial target test delta functions at 22.3, 22.9 and 25.6eV (peaks B, C and D′) have been used for the set of 80 Ta 4f spectra. Fig. 3 shows (dotted lines) the spectra of pure components obtained after ITTFA and scaled at constant area. In addition, the spectrum corresponding to the clean Ta surface, $Ta^0$, is also given for comparison. It should be indicated that, a similar spectrum is obtained by ITTFA using a test delta function at 21.7eV (peak A). However, in this case the pure $Ta^0$ spectrum was already experimentally available. Moreover, only a test delta function for every Ta doublet related to the different Ta-N phases has to be used in the ITTFA procedure, to predict both Ta $4f_{7/2}$ and $4f_{5/2}$ pure components, with a *sos* of ~1.9eV in good agreement with the values reported in the literature[6,16,25]. A fit of these pure component spectra with two asymmetric gaussian-lorentzian functions of the same full width at half maximum, *fwhm*, (dashed lines) gives a Ta $4f_{7/2}/4f_{5/2}$ area ratio, *R*, close to the theoretical value of 8/6, supporting the accuracy of the pure predicted spectra. The parameters characteristic of the calculated pure component spectra of Fig. 3 are given in Table 1.

Prieto et al[26] and Takano et al[17,27] have reported a value of ~22.85eV for the Ta $4f_{7/2}$ binding energy of a RSP $Ta_2N$ film and a $TaN_{0.28}$ film grown by 6keV $N_2^+$ implantation, respectively. According to the Ta–N phase diagram, for a $TaN_x$ compound, the range $0.28 \leq x \leq 0.54$ can be attributed to the hexagonal γ-$Ta_2N$ phase[7,18]. Therefore, the pure component spectrum obtained by ITTFA for peak C should be attributed to the hexagonal γ-$Ta_2N$ phase. On the other hand, the pure component spectra obtained by ITTFA for peaks B and D′ should correspond to tantalum nitride phases with lower and higher nitrogen content than γ-$Ta_2N$, respectively, because their binding energies, and therefore the charge transfer from tantalum to nitrogen, are lower and higher than that of the γ-$Ta_2N$ phase, respectively. In particular, the pure component at 22.3eV can be associated with the cubic β-$TaN_{0.05}$ phase, in good agreement with the value reported by Baba et al[27,28] for a $TaN_{0.07}$ film grown by 8keV $Ar^+$-ion bombardment of a Ta substrate in nitrogen atmosphere. This phase was first reported by Raole et al in nitrogen-implanted tantalum[13]. Zhang et al[3] have reported a value of ~23.5eV for the Ta $4f_{7/2}$ band of a film of cubic TaN formed by IBAD. However other possibilities as the hexagonal δ-TaN and ε-TaN phases cannot be ruled out.

It should be noted, that contrarily to the hexagonal TaN phases, the cubic TaN phase is not stable under thermodynamic equilibrium conditions. However, this phase has been reported by several authors in tantalum nitride films grown by ion beam based methods[2-4,13,15]. Therefore, it is not possible to distinguish between the hexagonal or cubic TaN phases from the XPS data of this work. Moreover, the *fwhm* of the pure component spectra obtained by ITTFA from peak D′ is greater than those obtained for the $Ta^0$, $\beta$-$TaN_{0.05}$ and $\gamma$-$Ta_2N$ phases (see table 1), suggesting the contribution of several phases to the spectrum. According to that, the band at 23.6eV could be attributed either to a hexagonal or a cubic TaN phase, or to a mixture of both, and will be denoted as cubic/hexagonal TaN.

Once accurate spectra for all the pure components are obtained by ITTFA, they are used to obtain their respective concentrations or weight factors in every experimental spectrum carrying out the Target Testing transformation with spectra[19,21-23]. Fig. 4 shows the evolution of the Ta 4f pure component concentrations as a function of the ion dose for the ion energies used in this work, obtained after Target Testing transformation with the pure component spectra of Fig. 3. Continuous lines in Fig. 3 show the reproduction of the target spectra (dotted lines) after TT transformation. As can be seen the agreement is very good, supporting again the validity of the predicted pure component spectra as representative of the principal factors. Moreover, an excellent reproduction of all experimental spectra is obtained making a linear combination of the pure component spectra weighted by their respective concentrations.

Several stages can be observed for the concentration evolutions of Fig. 4. In a first stage, a strong attenuation of the $Ta^0$ signal is accompanied by the formation of $\beta$-$TaN_{0.05}$ species. In this first stage, the nitrogen implanted atoms occupy octahedral interstitial positions in the bcc tantalum matrix, forming the $\beta$-$TaN_{0.05}$ phase with a slightly larger lattice parameter than that of tantalum[12,13]. In a second stage, as the ion dose increases and therefore more nitrogen is available, the transformation of β-TaN$_{0.05}$ into γ-Ta$_2$N and cubic/hexagonal TaN phases is observed, in good agreement with the XRD results of Zhou et al[12]. It is interesting to note, that the γ-Ta$_2$N signal reaches a maximum for intermediate ion doses. With increasing ion dose, the γ-Ta$_2$N signal slowly decreases due to its further transformation into cubic/hexagonal TaN phases to finally reach the saturation. During this second stage, a lower attenuation of the Ta$^0$ signal is observed. It is worth to mention that the ion doses at which β-TaN$_{0.05}$ and γ-Ta$_2$N phases reach their maximum and at which saturation is reached decrease with increasing ion beam energy. This is attributed to an increase of the reaction cross-section with decreasing the ion beam energy during nitrogen implantation[14]. At saturation, the tantalum nitride layers formed are a mixture of γ-Ta$_2$N and cubic/hexagonal TaN phases, the relative proportion being dependent on both the ion dose and ion energy. However, at this stage, small Ta$^0$ and β-TaN$_{0.05}$ signals can be observed, probably due to the ion-bombardment-induced reduction of the tantalum nitride previously formed or to the limited thickness of the film. The coexistence of Ta$_2$N and TaN phases found in this work is in good agreement with previous results[12,13].

Fig. 5 shows the nitrogen concentration depth profiles, $C_N(z)$, obtained using TRIDYN simulations[29,30] for nitrogen implantation in metallic tantalum at beam energies of 0.5, 1, 2, 3 and 5keV, using the same experimental saturation doses as in Fig. 4. Likewise, the sputtering yields at saturation for tantalum and nitrogen, $S_y^{Ta}$ and $S_y^N$, have been also obtained during the simulations and are given in Table 2. Monte Carlo TRIDYN code (version 4.0) accounts for the dynamic change of the thickness and composition of the target substrate during nitrogen implantation, as well as the changes in the nitrogen and tantalum sputtering yields, preferential sputtering and atomic mixing mechanisms. Tantalum nitride film thicknesses of ~ 59, 68, 78, 93 and 116Å, for 0.5, 1, 2, 3 and 5keV nitrogen implanted Ta substrates, respectively, are obtained from data of Fig. 5. Considering that the attenuation length of Ta 4f photoelectrons is estimated to be ~ 27Å[31], the small $Ta^0$ signal observed at saturation in Fig. 4 is probably due either to the limited thickness of the film or to the presence of metallic reduced species in it. Since the $Ta^0$ species disappear upon increasing the takeoff angle up to $\theta=70°$, i.e. with decreasing the probing depth, they are not located at the near surface. On the other hand, it should be noted that the low-stoichiometry $\beta$-$TaN_{0.05}$ phase completely disappears at saturation for ion beam energies $E_p \geq 2keV$ due to its transformation into more nitrogen-rich nitrides, whereas for 0.5 and 1keV unexpectedly reappears at higher ion doses. Raole et al[13] have suggested that preferential nitrogen loss during 30keV nitrogen implantation of tantalum is responsible for the formation of the low-stoichiometry $Ta_2N$ phase at higher ion doses (~$5 \times 10^{17}$ atoms/cm$^2$) due to the reduction of cubic TaN and hexagonal $Ta_5N_6$ phases previously formed. According to results of Table 2, the nitrogen sputtering yield, $S_y^N$, at saturation does not depend markedly on the ion beam energy between 0.5 and 5keV. However, the $S_y^N/S_y^{Ta}$ ratio increases strongly for $E_p < 2keV$. This behaviour could explain the existence of $\beta$-$TaN_{0.05}$ species at saturation for 0.5 and 1keV ion beam energies as due to preferential sputtering of nitrogen for intermediate and higher ion doses, in agreement with the Raole´s et al results[13]. However, we have not observed the above mentioned nitrogen loss effect for ion beam energies between 2 and 5keV because the $S_y^N/S_y^{Ta}$ ratio is ~1.

Considering that the attenuation length of N 1s photoelectrons is estimated to be $\lambda_N \approx 22$Å[31], the nitrogen concentration measured for a takeoff angle $\theta$ can be calculated by Eq. (1):

$$C_N(\theta) = \frac{1}{\lambda_N \cos\theta} \int_0^\infty C_N(z) \exp\left(\frac{-z}{\lambda_N \cos\theta}\right) dz \qquad (1)$$

The experimental nitrogen concentration, $C_N^{exp}$, has been calculated from the average composition, $x$, of the TaN$_x$ film formed as discussed in our previous work[14]. In Table 3, the $C_N^{exp}$ values are compared with those obtained by integration of $C_N(z)$ TRIDYN data according to Eq. (1) using the experimental takeoff angle, $\theta=0°$. The observed differences suggest that in addition to the $N_2^+$ implantation and atomic mixing mechanisms, simulated by the TRIDYN code, other mechanisms such as ion beam enhanced diffusion or chemical reaction with the nitrogen partial pressure of the ambient can also play an important role in the nitridation of the tantalum substrate, above all for higher ion beam energies, as previously suggested for hafnium[19]. Furthermore, Takano et al[17,28] have also suggested that the surface adsorption on metals reactive to nitrogen can be enhanced during implantation.

4. CONCLUSIONS

Tantalum nitride thin films have been formed by nitrogen implantation at room temperature of metallic tantalum in the energy range of 0.5−5keV. Quantitative information has been obtained by application of FA, comprising PCA and ITTFA, to the Ta 4f XPS spectra. The number, spectral shape and concentrations of the different Ta−N phases formed during nitrogen implantation have been obtained without any prior assumptions. The tantalum nitride films consist of a mixture of metallic tantalum, β-TaN$_{0.05}$, γ-Ta$_2$N and cubic/hexagonal TaN phases, their composition depending on both the ion dose and ion energy.

The kinetics of nitridation is characterised by two stages. In a first stage, a strong attenuation of the metallic $Ta^0$ signal is accompanied by the increase of $\beta$-$TaN_{0.05}$ species. In a second stage, $\beta$-$TaN_{0.05}$ species are transformed into $\gamma$-$Ta_2N$ and cubic/hexagonal TaN species. The concentration of $\gamma$-$Ta_2N$ species reaches a maximum for intermediate ion doses, slowly decreasing after that because of its further transformation into cubic/hexagonal TaN phases for higher ion doses up to saturation. At saturation, the tantalum nitride layers formed are a mixture of $\gamma$-$Ta_2N$ and cubic/hexagonal, but small $Ta^0$ and $\beta$-$TaN_{0.05}$ signals can be observed due to preferential sputtering of nitrogen and the limited thickness of the film.

Comparison of the experimental nitrogen concentration obtained from the average composition with that obtained from TRIDYN simulations suggests that in addition to nitrogen implantation and atomic mixing, other mechanisms like ion beam enhanced diffusion or the chemical reactivity of the tantalum substrate towards nitrogen should be also considered.

ACKNOWLEDGEMENTS

The authors thank D. Díaz for technical assistance. Likewise, the authors gratefully acknowledge Dr. M. Posselt (Institute of Ion Beam Physics and Materials Research. Forschungszentrum Rossendorf, Dresden, Germany) for the license of TRIDYN program. This work was financially supported by the Spanish Ministerio de Ciencia y Tecnología (projet MAT2002-04037-C03-02).

TABLES

Table 1: Parameters of the pure component spectra of Fig. 3.

|  | Binding energy (eV) | | sos(eV) | fwhm(eV) | R |
|---|---|---|---|---|---|
|  | $4f_{7/2}$ | $4f_{5/2}$ | | | |
| $Ta^0$ | 21.7 | 23.6 | 1.9 | 1.2 | 1.33 |
| (B)→ β-TaN$_{0.05}$ | 22.3 | 24.2 | 1.9 | 1.3 | 1.38 |
| (C)→ γ-Ta$_2$N | 22.9 | 24.8 | 1.9 | 1.3 | 1.36 |
| (D′)→ cubic/hexag. TaN | 23.6 | 25.5 | 1.9 | 1.6 | 1.34 |

Table 2: Tantalum and nitrogen sputtering yields, $S_y^{Ta}$ and $S_y^N$, at saturation for different ion beam energies, $E_p$, calculated using TRIDYN simulations.

| $E_p$ (keV) | $S_y^{Ta}$ | $S_y^N$ | $S_y^N/S_y^{Ta}$ |
|---|---|---|---|
| 0.5 | 0.04 | 0.46 | 11.50 |
| 1 | 0.14 | 0.57 | 4.07 |
| 2 | 0.40 | 0.41 | 1.02 |
| 3 | 0.51 | 0.48 | 0.94 |
| 5 | 0.60 | 0.55 | 0.92 |

Table 3: Comparison of nitrogen concentration, $C_N$, at saturation for different ion beam energies, $E_p$, calculated from experimental data and TRIDYN simulations of Fig. 5.

| $E_p$ (keV) | $C_N^{exp}$(%) | $C_N^{TRIDYN}$(%) |
|:---:|:---:|:---:|
| 0.5 | 47.8 | 52.7 |
| 1 | 47.6 | 46.5 |
| 2 | 41.6 | 29.9 |
| 3 | 39.2 | 30.5 |
| 5 | 36.9 | 30.2 |

FIGURE CAPTIONS

Fig. 1: Evolution of the Ta 4f core level spectrum of a Ta substrate for ion doses up to saturation after implantation at room temperature by 1 and 5keV $N_2^+$ ions.

Fig. 2: TTR profile for the Ta 4f set of 80 spectra involving all the ion doses and ion energies spectra measured in this work, in continuous line. Dashed line shows the TTR profile obtained for the same set of spectra after subtraction of the metallic $Ta^0$ contribution (subtraction is explained in the text).

Fig. 3: Ta 4f pure component spectra of the different Ta−N phases formed during nitrogen implantation obtained after ITTFA as explained in the text (dotted lines). Continuous lines show their reproduction after Target Testing transformation. Dashed lines represent the asymmetric gaussian-lorentzian functions used to calculate the Ta $4f_{7/2}/4f_{5/2}$ area ratio.

Fig. 4: Ta 4f pure component concentrations as a function of the ion dose for several ion energies.

Fig. 5: Nitrogen concentration depth profiles, $C_N(z)$, obtained using TRIDYN simulations of the nitrogen implantation in metallic tantalum at ion beam energies of 0.5, 1, 2, 3 and 5keV, using the experimental ion doses at saturation of Fig. 4.

Composition of tantalum nitride thin films formed by low-energy nitrogen implantation: a Factor Analysis study of the Ta 4f XPS core level; by A. Arranz and C. Palacio.

Fig. 1

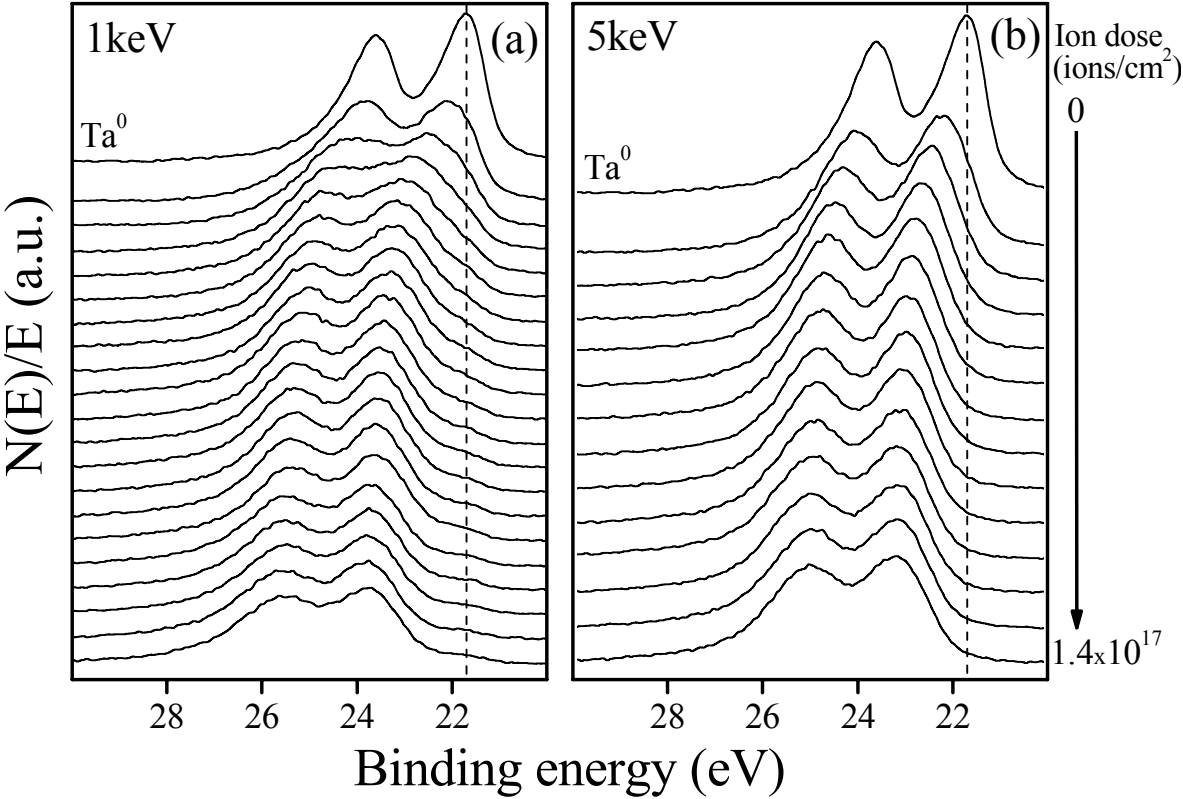

Composition of tantalum nitride thin films formed by low-energy nitrogen implantation: a Factor Analysis study of the Ta 4f XPS core level; by A. Arranz and C. Palacio.

Fig. 2

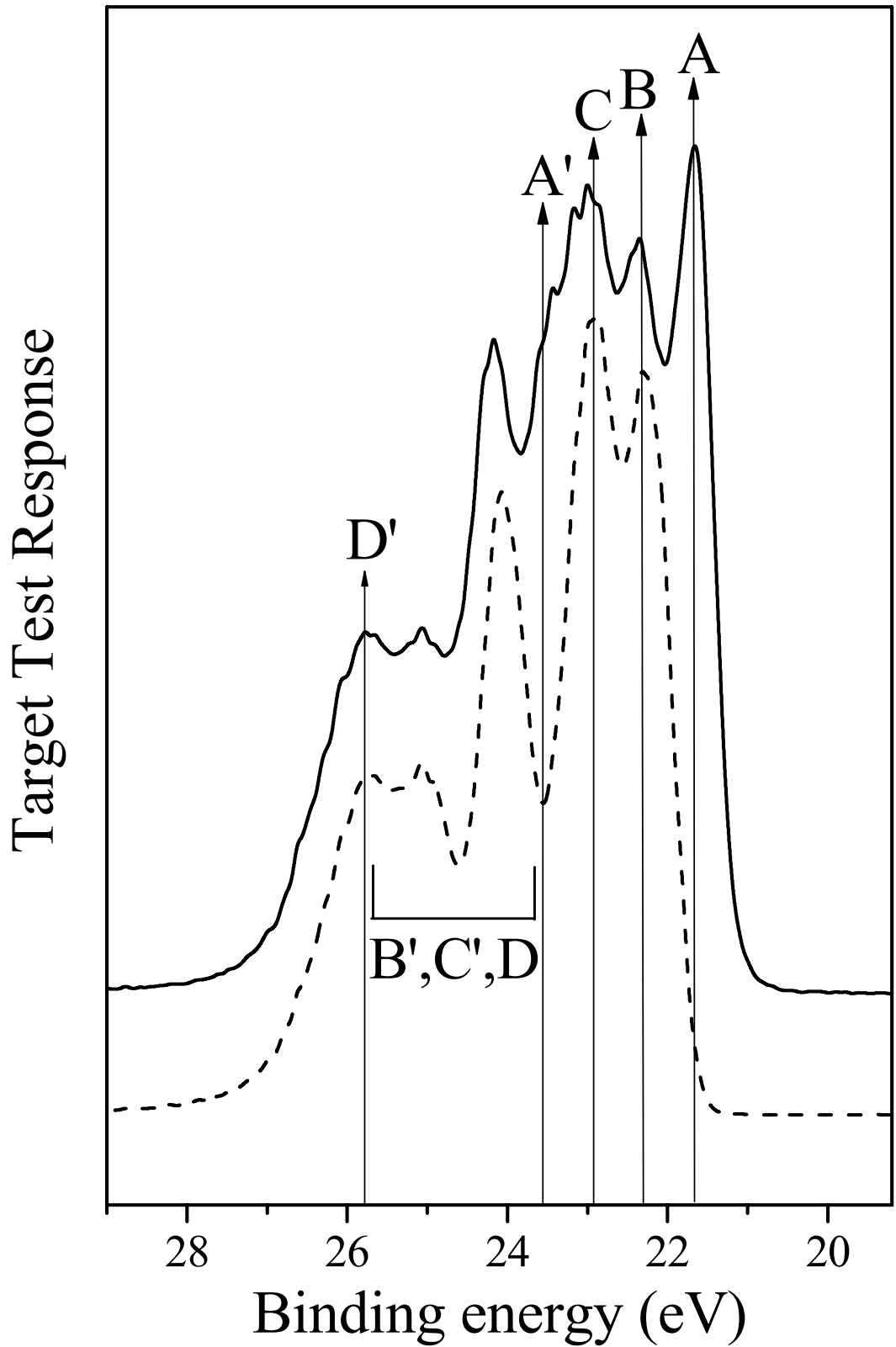

Composition of tantalum nitride thin films formed by low-energy nitrogen implantation: a Factor Analysis study of the Ta 4f XPS core level; by A. Arranz and C. Palacio.

Fig. 3

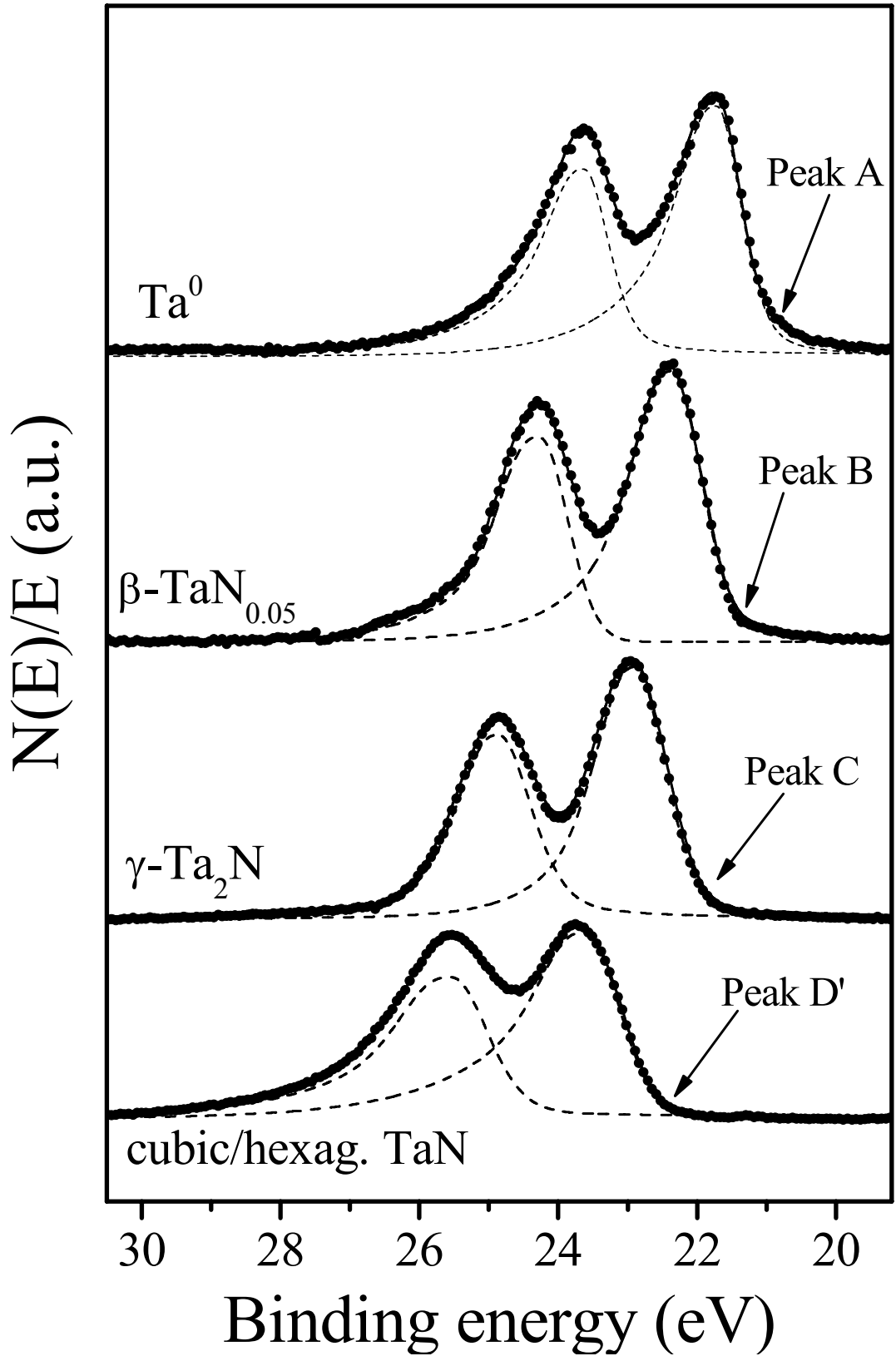

Composition of tantalum nitride thin films formed by low-energy nitrogen implantation: a Factor Analysis study of the Ta 4f XPS core level; by A. Arranz and C. Palacio.

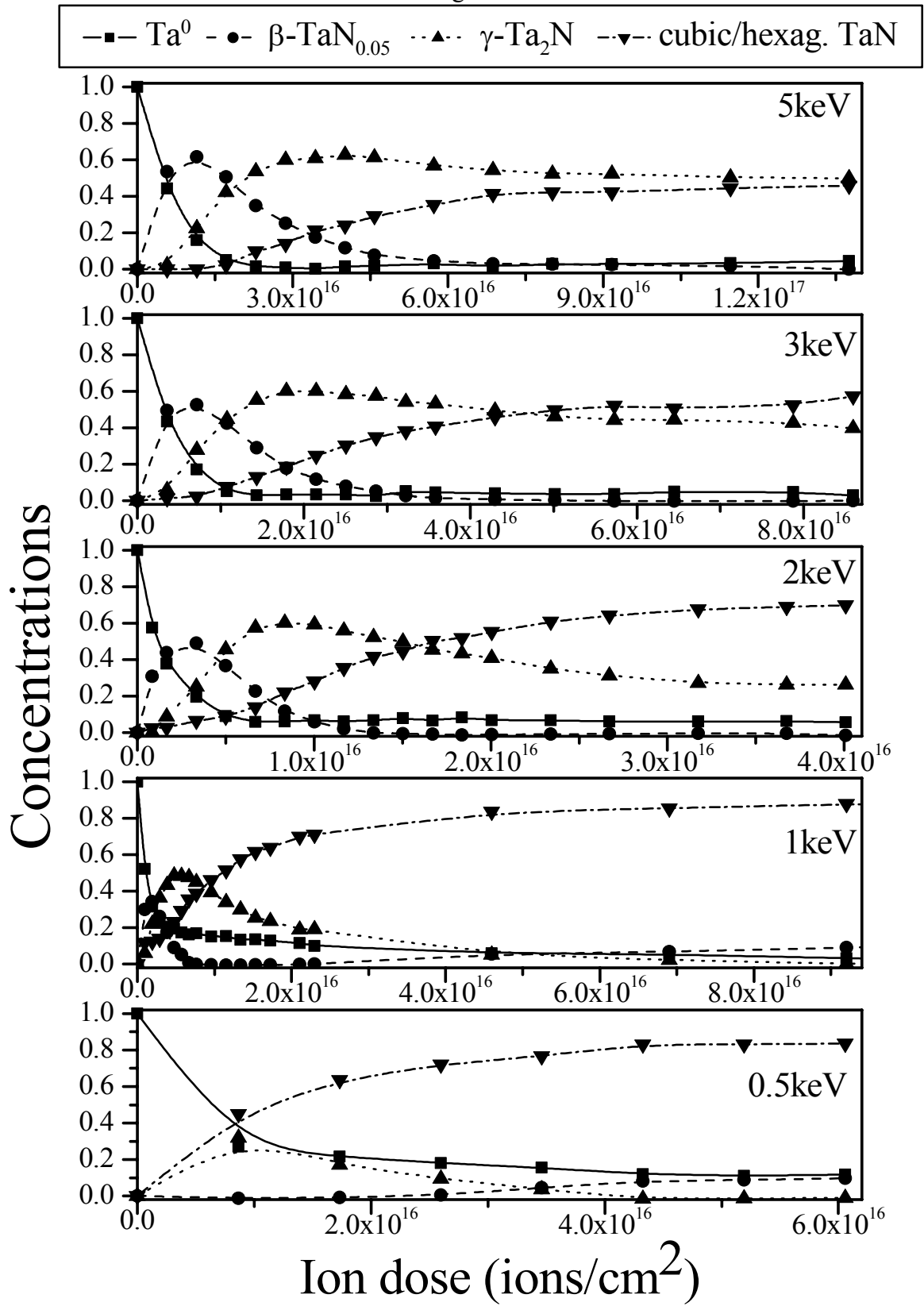

Fig. 4

Composition of tantalum nitride thin films formed by low-energy nitrogen implantation: a Factor Analysis study of the Ta 4f XPS core level; by A. Arranz and C. Palacio.

Fig. 5

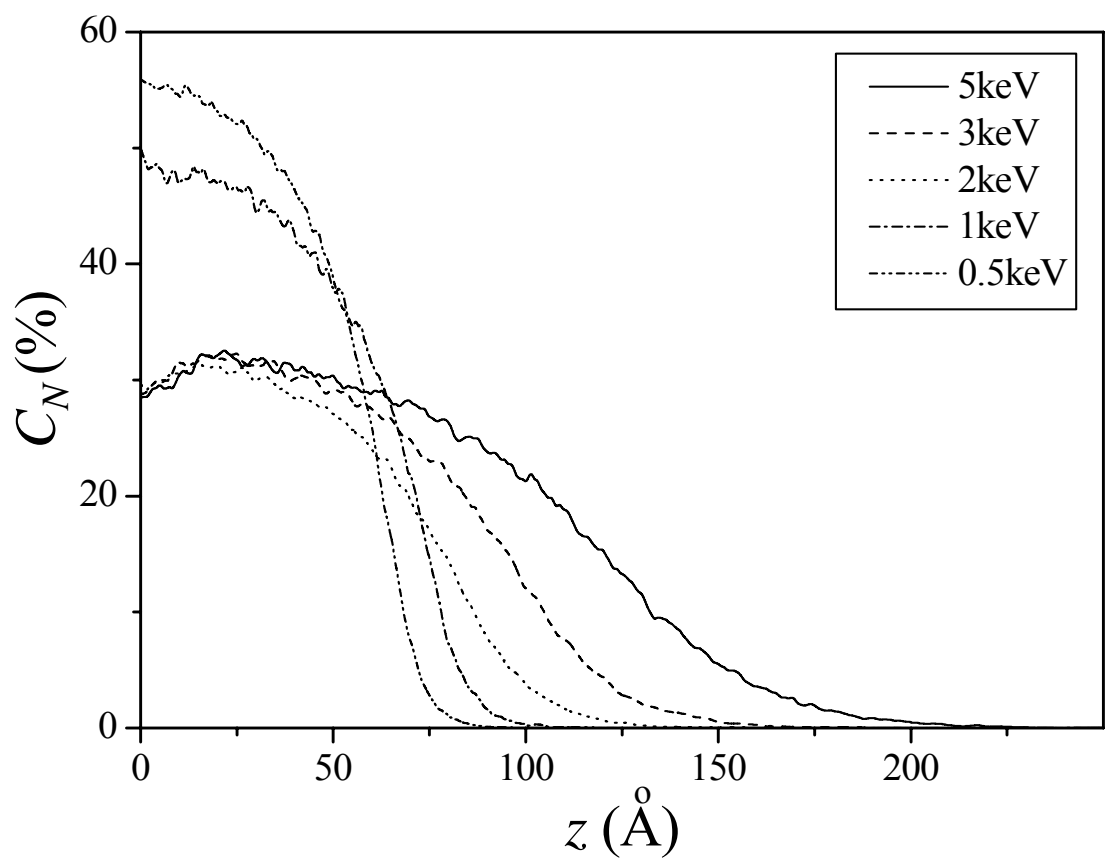